\documentclass[aps,prl,reprint,superscriptaddress]{revtex4-1}

\usepackage{graphicx,amssymb,amsmath}

\begin{document}


\title{Fluctuating Mobility Generation and Transport in Glasses}


\author{Apiwat Wisitsorasak}
\affiliation{Center for Theoretical Biological Physics, Rice University, Houston, TX 77005, USA}
\affiliation{Department of Physics \& Astronomy, Rice University, Houston, TX 77005, USA}

\author{Peter G. Wolynes}
\email[]{pwolynes@rice.edu}
\affiliation{Center for Theoretical Biological Physics, Rice University, Houston, TX 77005, USA}
\affiliation{Department of Physics \& Astronomy, Rice University, Houston, TX 77005, USA}
\affiliation{Department of Chemistry, Rice University, Houston, TX 77005, USA}



\date{\today}

\begin{abstract}
In the context of the random first order transition theory we use an extended mode coupling theory of the glass transition that includes activated events to account for spatiotemporal structures in aging and rejuvenating glasses. We numerically solve fluctuating dynamical equations for mobility and fictive temperature fields which capture both mobility generation through activated events and facilitation effects. Upon rejuvenating, a source of high mobility at a glass surface initiates a growth front of mobility which propagates into the unstable low mobility region. The speed of the front is consistent with   experiments on the rejuvenation of stable glasses, which ``melt'' from their surface.
\end{abstract}


\maketitle

On human length and time scales glasses seem static, yet glasses constantly move at molecular scales. Molecules in glasses change their locations through rare, activated events at a rate which varies throughout the glass owing to its aperiodic structure. Motions in one location cause or relieve constraints thereby changing the rate of transitions in neighboring regions. From these considerations it follows that the apparent mobility of regions fluctuates both in space and time and appears to be locally transported. In this paper, we explore, using numerical methods, fluctuating mobility generation and transport in glasses focussing on rejuvenation after heating. The framework used is based on the random first order transition (RFOT) theory of glasses \cite{stoessel:4502, singh1985hard, kirkpatrick1987connections, kirkpatrick1987stable, xia2001microscopic, lubchenko2004theory, lubchenko2007theory}. 

The static appearance of glasses has been formally addressed by mean field theories such as the explicitly dynamical mode coupling theory (MCT) \cite{kirkpatrick1987connections,gotze2009complex} as well as by outwardly static approaches \cite{stoessel:4502,singh1985hard}. The similarities between these different approaches lead to the suggestion that they are approximately equivalent \cite{kirkpatrick1987connections}. These disparate analogous approaches indeed do provide exactly equivalent results for a large class of spin glass models with infinite range interactions \cite{kirkpatrick1987stable,PhysRevLett.58.2091,Bouchaud1996243}. In mean field models, a dynamic transition to immobility occurs discontinuously at a high dynamic transition temperature, $T_d$, below which an exponentially large number of frozen states emerges. At a lower temperature, $T_K$, the configurational entropy of these states vanishes and a true thermodynamic transition ensues \cite{kirkpatrick1987stable}. For systems with finite range interactions, the transition at $T_d$ is wiped out by local activated transitions between the frozen states \cite{kirkpatrick1987stable}. RFOT theory argues that activated transitions resembling nucleation events become the source of mobility in deeply supercooled liquids. These thermally activated motions would cease at the thermodynamic transition if the entropy crisis occurs or otherwise only at absolute zero.

While the activated mechanism of mobility generation is a key aspect of RFOT theory,  owing to its connection with the mode coupling theory, RFOT theory shares with kinetically constrained models \cite{SollichConstrained2003} the idea that mobility can be transported i.e. the notion of facilitation. Mobility transport effect is an essential feature of RFOT theory, diminishing the effects of the instantaneous heterogeneity in activation barriers which otherwise would lead to a broader distribution of relaxation times than observed \cite{xia2001microscopic}.

The idea that the mobility of a region of the fluid depends on the mobility of its neighbors provides the historic basis for MCT.  Bhattacharyya, Bagchi and Wolynes (BBW) suggested a way to bridge the gap between MCT which is adequate above the dynamical transition and mobility generation through activated transitions which occurs below the mean field dynamic transition by including an activated event vertex in the self-consistent MCT \cite{bhattacharyya2005bridging,bhattacharyya2008facilitation}. BBW's treatment did not make explicit the spatiotemporal nature of the dynamical heterogeneity. To make clearer the connection we can allow the mobility to be an explicitly varying dynamical variable in the BBW analysis. The resulting coupled inhomogeneous nonlinear self-consistent differential-integral equations are  cumbersome to study but assuming slow variations in time and space in the mobility allows a continuum approximation to be made \cite{biroli2006inhomogeneous,wolynes2009spatiotemporal,stevenson2008surface}. This gives a simple description of a rich set of phenomena especially in the case of ``rejuvenating'' glasses that are heated after falling out of equilibrium. Wolynes argued on the basis of the continuum theory that mobility transport and generation conspire to set up moving fronts in heated glasses that resemble flames \cite{wolynes2009spatiotemporal}. Ignition by mobility generation occurs at points within the bulk but cooperative rearrangements occur with especially large rapidity at a free interface which acts as an external source for a rejuvenation front  \cite{stevenson2008surface}. The theory thus predicts then there are both heterogeneous and homogeneous mechanisms by which glass ``melts'' into an equilibrated fluid.  These two mechanisms have been observed by Ediger in his studies of ultrastable glasses prepared by vapor deposition \cite{swallen2010transformation}. The approximate continuum treatment of rejuvenation agrees qualitatively with the more precise later measurements \cite{swallen2010transformation,sepulveda:204508,kearns2010observation} but according to Ref.~\onlinecite{sepulveda:204508} the simple analytical result for the front speed is quantitatively off by about two orders of magnitude. In the first set of measurements \cite{swallen2010transformation} there was also evidence for a modest overshoot phenomenon in which, just after a front passes, the mobility seems to be higher than it will eventually become in the fully equilibrated liquid. This overshoot, not anticipated by Wolynes's calculation, is, however, not apparent in later experimental studies \cite{sepulveda:204508}. The possibility of a large deviation from the simple approximation inspired us to explore more completely the continuum equations for mobility generation and transport in glasses. Here we show that a more accurate numerical treatment of the fluctuating continuum equations gives better quantitative agreement with observed front propagation speed but no sign of the overshoot occurs in the average behavior. 

We first recount the continuum equations emphasizing the role of fluctuations. Starting from the spatially inhomogeneous MCT with activated events we approximate the memory kernel as being explicitly time and space dependent but treat it as locally frequency independent thus representing a mobility field $\mu(x,t)$. Expanding the inhomogeneous equation in Taylor series yields a continuum equation which we can supplement with local random terms \cite{wolynes2009spatiotemporal}
\begin{equation}\label{eq:LIMCTMuNoFluc}
\frac{\partial \mu}{\partial t}  = \frac{\partial }{\partial x}  \left( \bar \mu\xi^2 \frac{\partial \mu}{\partial x}  \right) - \bar \mu \left(\mu - \bar\mu \right) + \delta g + \frac{ \partial \delta j}{\partial x}
\end{equation}
where $\bar \mu$ is the uniform solution of the homogeneous MCT equation and $\xi$ is the size of a cooperatively rearranging region (CRR). We will refer to Eq.~(\ref{eq:LIMCTMuNoFluc}) as linearized inhomogeneous mode coupling theory (Li-IMCT). In deeply glassy systems $\bar\mu$ the mobility from the uniform solution of the BBW equations can be reasonably approximated by the traditional aging formulation of Narayanaswamy-Moynihan and Tool in term of a fictive temperature $T_F$, which measures the local energy, as in RFOT theory by Lubchenko and Wolynes \cite{lubchenko2004theory} $ \bar\mu(T_F,T) = \mu_0 \exp \left\{- \frac{x E^\ddag}{k_B T} - \frac{\left(1-x \right) E^\ddag}{k_B T_F} \right\}$.
The parameter $x$ is predicted by RFOT theory to be inversely proportional to the heat capacity discontinuity $\Delta c_p$.  To complete the description we take the local fictive temperature to approach the ambient temperature through an ultralocal relaxation law
\begin{equation}\label{eq:TF}
\frac{\partial T_F}{\partial t} =  - \mu \left( T_F - T \right) + \delta \eta.
\end{equation}

Since activated motions occurring right at the surface feel no mismatch energy penalty on the interface along the free surface they have to overcome only the mismatch energy on the hemisphere facing the bulk \cite{stevenson2008surface}. Free surfaces thus represent a  source of inhomogeneous mobility. In the bulk itself there are dynamic inhomogeneities from fluctuations in the local fictive temperature and from the activated events which transport the mobility field. We  treat the random force terms as coarse-grained white noises with correlation lengths reflecting the length scale of  the activated events. The intensities of the noises may be found by requiring the linearized equations to satisfy locally the fluctuation-dissipation relations. The local fictive temperature fluctuation $\delta \eta$ is taken to be $\label{eq:TfNoiseLIMCT} \langle \eta(x,t) \eta(x',t') \rangle = 2 \mu T^2 \frac{k_B}{\Delta c_p N^\ddag} \delta(x-x') \delta(t-t') $,
where $N^\ddag$ is the number of molecular units in a CRR. The fluctuations in mobility generation $\delta g$ and transport $\delta j$ arise due to fluctuations in free energy barrier heights which  are the cause of stretched exponential relaxation with a bare stretching parameter $ \beta_0 = 1/\sqrt{1+(\delta F^\ddag/k_B T)^2}$ \cite{xia2001microscopic}.
Linearizing the equations and treating $\bar \mu$ as constant, the fluctuation-dissipation relation yields a mobility generating noise  $\langle \delta g(x,t) \delta g(x',t') \rangle =  2 \bar \mu \mu^2 \left( \frac{1}{\beta_0^2} - 1\right) \left(\frac{T_g}{T} \right)^2 \delta (x-x') \delta(t-t') \nonumber$ and a mobility random flux  $\langle \delta j(x,t) \delta j(x',t') \rangle =  2 \bar  \mu \xi^2 \mu^2 \left( \frac{1}{\beta_0^2} - 1\right) \left(\frac{T_g}{T} \right)^2 
\delta (x-x') \delta(t-t') $ \cite{eyink1996hydrodynamics}.

While Li-IMCT is a natural perturbation result, it is worth noting that to the same order of perturbation the rate of relaxation of $\mu$ to its local value could be taken as $\mu$ itself rather than $\bar\mu$. This leads to a nonlinear transport equation $ \label{eq:KPPMCTMuNoNoise}
\frac{\partial \mu}{\partial t}  = \frac{\partial }{\partial x}  \left( \mu\xi^2 \frac{\partial \mu}{\partial x}  \right) -  \mu \left(\mu - \bar\mu \right)  + \delta g + \frac{\partial \delta j}{\partial x} $. This equation has the structure of the Fisher-Kolmogorov-Petrovsky-Piskounov equation \cite{KPP1937,fisher1937wave} so we call this set of coupled equations the FKPP inhomogeneous mode coupling theory (FKPP-IMCT). This equation turns out to do a somewhat better job of describing the experiments than the Li-IMCT does.

Eq.~(\ref{eq:LIMCTMuNoFluc}) resembles the diffusion equation for some kind of excitation. Of course, such ``excitations'' are quite fictional -- there is no regular structure in a glass that can be used to a priori define a defect. The formal analogy between inhomogeneous mode coupling and excitation models explains why such models while being microscopically unrealistic do manifest phenomena reminiscent of molecular glasses.

Exact analytical treatment of the equations for mobility generation and transport is difficult. It has been argued, however, that for aging glasses i.e. systems that have been quickly quenched to low temperatures, the  aging dynamics that can be easily studied in the laboratory will not be very different from what occurs initially and so involves primarily the mobility generation events themselves \cite{wolynes2009spatiotemporal}. When a glass is heated however the autocatalytic interplay between mobility generation and transport qualitatively changes the picture. Recognizing the analogy to combustion, Wolynes invoked some of the simpler approximations used in the analysis of flames \cite{barenblatt1971intermediate} which predict the front propagates at a speed $\sqrt{2/3} \xi \mu^{\mathrm{high}}$. This upper bound to the speed depends only on the mobility in the high $T$ equilibrated material, $\mu^{\mathrm{high}}$, but is independent of the mobility of the initially prepared low mobility material. Here we test that idea by numerical methods. 

The numerical calculations were carried out using parameters referenced to tris(naphthyl)benzene or TNB \cite{swallen2010transformation} whose glass transition temperature for $\tau = 100\ s$, $T_g$ is $347\ K$, with $\xi$ = 2.5 nm. In 1D, the calculations use a lattice of 200 cells with a spacing $\Delta x = 1\ $nm. For the  2D case, a square two-dimensional lattice of $20\times100$ cells with mesh size $\Delta x = 1\ $nm was employed. Periodic boundary condition are assumed for all cases but the surface is represented by a region with intrinsically high mobility given by the Stevenson-Wolynes formula $\tau_s = \sqrt{\tau_0 \tau_\alpha}$ \cite{stevenson2008surface}. Owing to the stiffness of the equations we cannot study them at as low a fictive temperature as the experiment. 

We first report numerical results obtained from 1D deterministic theory. Due to the symmetry of the deterministic problem, these also apply in two dimensions. In the Li-IMCT case, the interface width as measured by full-width-at-half-maximum of the source term $-\bar \mu (\mu - \bar \mu)$ is about 5 nm. This scale is only a bit larger than the size of the entropic droplets that generate mobility \cite{xia2001microscopic}. The width of the interface from the FKPP-IMCT is as wide as 10 nm, more consistent with the experimental measurement \cite{swallen2010transformation}. 

\begin{figure}
  \includegraphics[width=19pc,angle=0]{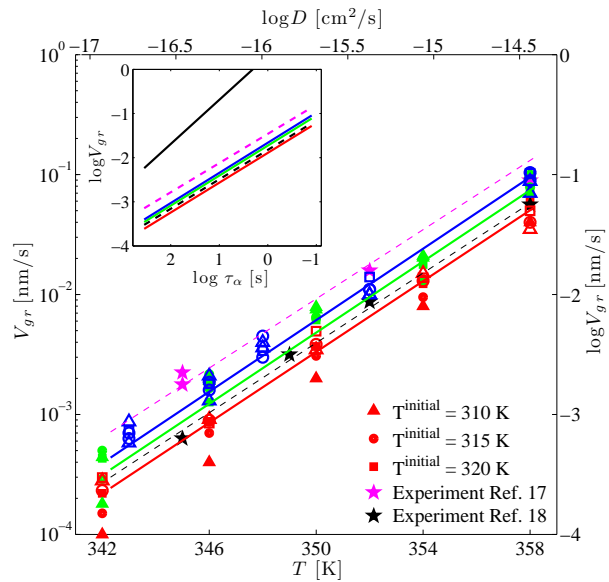}
  \caption{ (Color online) The speed of the mobility front from the dynamics of the 1D deterministic Li-IMCT (solid red markers), the 1D deterministic FKPP-IMCT (opened red markers), the 1D stochastic Li-IMCT (solid green markers), the 1D stochastic FKPP-IMCT (opened green markers), the 2D stochastic Li-IMCT (solid blue markers), and the 2D stochastic FKPP-IMCT (opened blue markers). Different kinds of symbols indicate different initial fictive temperatures, the triangles are for $T^{\mathrm{initial}} = 310 K$, circles for $T^{\mathrm{initial}} = 315 K$, and squares for $T^{\mathrm{initial}} = 320 K$. The experimental measurements as reported in Ref.~\onlinecite{swallen2010transformation} are shown as magenta stars and the black stars represent the more recent results from Ref.~\onlinecite{sepulveda:204508}. All lines are drawn as guides for the eye. Inset: The front speed versus relaxation time. The continuum approximation of the front speed $\sqrt{2/3} \xi \mu^{\mathrm{high}}$ is drawn as the black solid line.  Notice the front speed does not scale quite as strongly as $\tau_\alpha^{-1}$. }\label{fig:f2}
\end{figure}

The speeds of front propagation at various temperatures predicted from the 1D deterministic Li-IMCT and FKPP-IMCT are shown in Fig.~\ref{fig:f2}. Increasing the ambient temperature increases the speed.  The predicted linear trend of $\log(v_{gr})$ with temperature $T$ agrees with observations. No calculated point differs from the experimental line \cite{sepulveda:204508} by more than a factor of 2. The predicted speed from the deterministic Li-IMCT however does vary somewhat more with the initial fictive temperature in the simulation than in experiment. Wolynes's approximate solution on the other hand showed independence of the speed on the initial fictive temperature in agreement with observation. The discrepancy in the numerical study contrasting with the correct prediction from the approximation may arise because the initial fictive temperature  in the study is not low enough thus leading to a  ``cold boundary difficulty", a problem extensively discussed in combustion theory using intermediate asymptotics \cite{barenblatt1971intermediate}. 

\begin{figure}
	\includegraphics[width=19pc,angle=0]{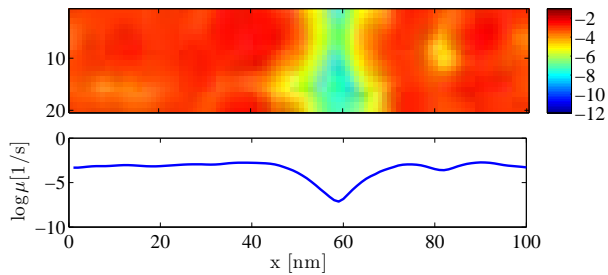}
	 \caption{(Color online) Mobility field snapshot at $t = 4.6 \times 10^4$ s. The upper plot shows the mobility field from 2D stochastic Li-IMCT. The color scheme denotes the mobility on a log-scale with corresponding color bars on the right. The figure in the lower panel show average values of the mobility fields along the $y$-axis.}\label{fig:f15}
\end{figure}

We now consider the stochastic models in one and two dimensions. While the three dimensional situation is numerically too demanding for us to treat presently the lower dimensional cases already allow a good comparison with experiment.  A realization of the front evolution from the 2D stochastic version of the Li-IMCT are shown in Fig.~\ref{fig:f15} through the mobility field $\mu$. Initially the sample is prepared in an ultrastable phase by simulating the stochastic equations for a uniform system at a low temperature, $T^{\mathrm{initial}} = 310\ K$.  The system is then instantaneously heated to a higher ambient temperature, $T = 342\ K$, making the former low mobility state unstable. As the fictive temperature increases, the front propagates into the bulk leaving the stable state behind. Due to the periodic boundary conditions fronts initiate at both ends and travel towards the ``inner'' bulk. 

In addition to the fronts emanating from the surfaces at high mobility reconfiguration events take place at some bulk locations faster than they do at others. Once these initially initiated nucleated regions reconfigure, they catalyze rearrangement of their low mobility neighbors and lead to fronts emanating radially from the rejuvenated centers -  as described earlier \cite{wolynes2009spatiotemporal}. Fast regions cannot as easily jump over slow regions which are more stable, so they often detour to avoid an immobile region or become jammed up at the interface between the two regions. This latter effect in a specific realization can give the appearance of the overshoot of the mobility behind the propagating front as shown in Fig.~\ref{fig:f15}. Both stochastic Li-IMCT and FKPP-IMCT equations give similar fluctuating overshoot phenomena.

To assign a typical speed to the stochastic front, the front position may be defined using a one dimensional integral of a threshold function $ z(t) = \frac{1}{2}\int_L dx \Theta[\mu(x,t) - \mu^{\mathrm{threshold}}] $ ,where $\Theta[f(x)]$ is Heaviside step function, $\mu^{\mathrm{threshold}}$ is a constant threshold \cite{garc'a1999noise}. The integration measures the amount of sample which has not yet been transformed much as in calorimetry \cite{kearns2010observation}. Physically $z(t)$ gives the depth of the front from each surface. The growth front speed is determined by $ v = \Delta z / \Delta t$, where $\Delta t$ is somewhat larger than a typical relaxation time $\mu^{-1}$. Averaging this quantity over initial times and initial ensemble yields the average front speed $v_{gr}$. 

The front speeds for ultrastable TNB glasses prepared at various initial temperatures from both 1D and 2D of stochastic Li-IMCT and FKPP-IMCT are displayed in Fig.~\ref{fig:f2}. The resulting speeds do slightly depend on the initial temperatures, but depend exponentially on the ambient temperature rapidly rising as the ambient temperature increases. The front speeds for all stochastic cases  are larger than for the deterministic cases  by a factor of two. These speeds from the stochastic simulations for both dynamics now deviate by less than 50\% from the front speeds  reported in Ref.~\onlinecite{sepulveda:204508}. The growth front velocities reported in Ref.~\onlinecite{swallen2010transformation} are about three times greater than those found in Ref.~\onlinecite{sepulveda:204508} in which it was suggested the samples prepared earlier may contain impurities such as the vacuum pump oil which increase the mobility. 

In the present paper we have numerically solved fluctuating continuum equations coming from  inhomogeneous MCT that were previously introduced to account for spatiotemporal structures in aging and rejuvenating glasses. Numerically accounting for fluctuations in mobility field and fictive temperature gives closer quantitative agreement with experimental measurements than the previous approximate treatment. 

The front propagation phenomena discussed here are of a very general nature. Any external influence that can increase mobility, if it is inhomogeneous in space can lead to such fronts. For example plasticizer diluents which lower $T_g$ will give fronts which will propagate into the bulk. The coupled diffusion may give an overshoot phenomenon like that seen in some of the experiments \cite{Rossi1995Phenomenological}. Stress also acts to increase mobility by lowering activation barriers \cite{Wisitsorasak17092012}. Shear banding \cite{Hays2000Microstructure} may arise from the coupling of stress to inhomogeneously propagating internal mobility fronts.

\begin{acknowledgments}
Financial support by the D.R. Bullard-Welch Chair at Rice University and a Royal Thai Government Scholarship to A.W. are gratefully acknowledged.
\end{acknowledgments}


\end{document}